# Quantum-fluid dynamics of microcavity polaritons

A. Amo[1], D. Sanvitto[1], D. Ballarini[1], F.P. Laussy[2], E. del Valle[2], M.D. Martin[1], A. Lemaître[3], J. Bloch[3], D.N. Krizhanovskii[4], M.S. Skolnick[4], C. Tejedor[2] & L. Viña[1]

[1]Dep. Física de Materiales, Univ. Autonóma de Madrid, 28049 Madrid, Spain

[2]Dep. Física Teórica de la Materia Condensada, Univ. Autonóma de Madrid, 28049 Madrid, Spain.

[3]LPN/CNRS, Route de Nozay, 91460, Marcoussis, France

[4]Dep. Physics & Astronomy, Univ.. of Sheffield, S3 7RH, Sheffield, U.K.

**Semiconductor microcavities offer a unique system to investigate the physics of weakly interacting bosons. Their elementary excitations, polaritons—a mixture of excitons and photons—behave, in the low density limit, as bosons that can undergo a phase transition to a regime characterised by long range coherence[1,2]. Condensates of polaritons have been advocated as candidates for superfluidity[3]; and the formation of vortices[4] as well as elementary excitations with a linear dispersion[5] are actively sought after. In this work, we have created and set in motion a macroscopically degenerate state of polaritons and let it collide with a variety of defects present in the sample. Our experiments show striking manifestations of a coherent light-matter packet that displays features of a superfluid, although one of a highly unusual character as it involves an out-of-equilibrium dissipative system where it travels at ultra-fast velocity of the order of 1% the speed of light. Our main results are the observation of i) a linear polariton dispersion accompanied with diffusion-less motion, ii) flow without resistance when crossing an obstacle, iii) suppression of Rayleigh scattering and iv) splitting into two fluids when the size of the obstacle is comparable with the size of the wavepacket. This work opens the way to the investigation of new phenomenology of out-of-equilibrium condensates.**

Below a critical temperature, a sufficiently high density of bosons undergoes Bose-Einstein condensation (BEC). Under this condition, the particles collapse into a macroscopic condensate with a common phase, showing collective quantum behaviour like superfluidity, quantised vortices, interferences, etc. Up to recently, BEC was only observed for diluted atomic gases at μK temperatures. Following the recent observations of non-equilibrium BEC in semiconductor microcavities at temperatures of ~10 K, using momentum-[1] and real-space[2] trapping, the quest is now towards the observation of the superfluid motion of a polariton BEC. For the same reasons that polaritons benefit from unusually favourable features for condensation, such as very high critical temperatures, it is expected that their superfluid properties would likewise manifest with altogether different magnitudes, such as very high critical velocities. Since they have shown many deviations in their Bose-condensed phase from the cold atoms paradigm, it is not clear a priori to which extent their superfluid properties would coincide or depart from those observed with atoms, among which quantised vortices[6], frictionless motion[7], linear dispersion for the elementary excitations[8], or more recently Čerenkov emission of a condensate flowing at supersonic velocities[9], are among the clearest signatures of quantum fluid propagation.

Microcavity polaritons are two-dimensional bosons of mixed electronic and photonic nature, formed by the strong coupling of excitons—confined in semiconductor quantum wells—with photons trapped in a micron scale resonant cavity. First observed in 1992[10], these particles have been profusely studied in the last fifteen years due to their unique features. Thanks to their photon fraction, polaritons can easily be excited by an external laser source and detected by light emission in the direction perpendicular to the cavity plane. However, as opposed to photons, they experience strong inter-particle interactions owing to their partially electronic fraction. Due to the deep polariton dispersion, the effective mass of these particles is $10^4$-$10^5$ smaller than the free electron mass, resulting in a very low density of states. This allows for a high state occupancy even at relatively low excitation intensities. However, polaritons live only a few $10^{-12}$ s in a cavity before escaping and therefore thermal equilibrium is never achieved. In this respect, a macroscopically degenerate state of polaritons departs strongly from an atomic Bose-condensed phase. The experimental observations of spectral and momentum narrowing, spatial coherence and long range order—which have been used as evidence for polariton Bose-Einstein condensation—are also present in a pure photonic laser[11]. The recent observation of long range spatial coherence[12], vortices[4] and the loss of coherence with increasing density in the condensed phase[13,14], are in accordance with macroscopic phenomena proper of interacting, coherent bosons[15]. But a direct manifestation of superfluidity is still missing.

In this work, using a new combination of temporally and spatially resolved spectroscopic techniques, we are able to probe directly the dynamics of motion of a non-equilibrium, coherent polariton quantum fluid, by tracking its space-time evolution. We report a shape preserving, non-diffusive propagation of matter dressed by the light field moving at a speed of the order of $10^6$ m s$^{-1}$. Studying the collision against obstacles of different sizes, we investigate the analogies of this state of matter with conventional BEC such as those realised with atomic gases, whose dynamics are known to display striking properties like superfluidity. We show that the study and manipulation of the condensate properties can thus be achieved in semiconductor chips of micrometer scale.

To make polaritons flow, we have to address three important issues: 1) the very short polariton dwell time in the cavity (< 2-5 ps), which hinders detection of their dynamics; 2) creation of polaritons with well-defined momentum; 3) spatial inhomogeneities given by sample strain and/or defects. Our experiments are based on the continuous replenishing of the polariton fluid, at energy $E_S$ and momentum $k_S$, from a higher-lying state, at $E_P$ and $k_P$, driven coherently by an external CW laser and ignited by a short trigger pulse (2 ps) at the idler state, $E_I$ and $k_I$, in a configuration of a triggered optical parametric oscillator (TOPO; see Fig. 1 and the Supplementary Information)[16]. The signal polaritons, fed by the pump in resonance with the lower polariton branch, last for more than $10^{-9}$ s after the trigger pulse of 2 ps, thus allowing the detection of the polariton flow dynamics. This approach solves the first issue listed above: although the lifetime of a single particle of the polariton droplet remains short (few picoseconds), the packet itself lasts hundreds of picoseconds. To reduce the effect of the spatial inhomogeneities present in all semiconductor microcavities[17,18], we keep the pump power at sufficiently high intensity so that most of the potential fluctuations are smoothed out[19]. Nonetheless, scattering centres are still present with an approximate density of $10^{-2}$ $\mu m^{-2}$ but, as shown below, they can be beneficial to reveal the peculiar quantum nature of the polariton fluid.

Images of the near- and the far-field of a GaAs based microcavity[20] are projected on to the entrance slit of a 0.5 m imaging spectrometer attached to a streak camera, thus allowing for simultaneous collection of two dimensional images, at a given energy, and resolved in time. To avoid problems caused by detection at the pulsed laser energy, which would bleach the streak camera, the pulsed laser is resonant with the idler rather than the signal state of the triggered OPO. Furthermore, to retain only the light emission from the TOPO signal, the OPO-only emission coming from the pump field, contributing at $k$=0, is subtracted from all the images. We can thus study the propagation of the signal polaritons which can have any, specifically selected, in-plane momentum $k$-vector given by the phase matching conditions shown in the caption of Fig. 1.

The experimental dispersions of the light emitted by the polaritons in different regimes are depicted in Fig. 2a. The false-colour graph in the upper panel presents the emission of the polariton fluid generated by the TOPO, after the trigger pulse has disappeared, and it is compared to the standard *parabolic* dispersion of the photoluminescence obtained under non resonant low-power excitation (lower panel). The most striking feature, under TOPO conditions, is the clear *linearization* of the dispersion around the signal state. It is also seen that, for the TOPO, polaritons are blue shifted ($\Delta E = 0.6$ meV at $k$=0), due to polariton-polariton interactions, and that the intensity peaks strongly at $k_S = +0.6$ μm$^{-1}$, which is determined by the phase-matching conditions ($2k_P = k_S + k_I$).

Our system involves macroscopically degenerate states of bosons coupled through exciton-exciton interactions and so can be accurately described theoretically by a nonlinear Schrödinger equation for the polariton wavefunction in the presence of a continuous pump $F_P$ and a pulse $F_I$, shining on the microcavity at an angle $k_{P,I}$, with frequency $\omega_{P,I}$ and localised in a Gaussian spot of spatial extension $\sigma_{P,I}$:

$$i\hbar\partial_t\psi(x,t) = \left[D - i\gamma/2 + V\left|\psi(x,t)\right|^2\right]\psi(x,t) + F_P e^{-(x-x_P)^2/\sigma_P} e^{i(k_P x - \omega_P t)}$$
$$+ F_I e^{-(x-x_I)^2/\sigma_I} e^{-(t-t_I)^2/\sigma_t} e^{i(k_I x - \omega_I t)}$$

(1)

where the operator $D$ is the free-propagation energy of the particles, i.e., in our case, provides the dispersion relation specific to polariton branches. The lower branch gives rise to the OPO or TOPO physics of phased-matched scattering. The two last terms are responsible for maintaining the system out of equilibrium against the decay $\gamma$. This represents a crucial difference with the usual case of atomic condensates. Equation (1) is integrated numerically, first in the absence of the pulse ($F_I$=0), until a steady state is reached for $|\psi|^2$. We obtain the energy-$k$ density plot of the system by Fourier-transforming $\psi(x,t)$. At a given time $t_I$, we release a Gaussian pulse ($F_I\neq0$) that triggers OPO processes. We track the new evolution of $\psi(x,t)$ until the steady state is restored, and Fourier-transform again during this time. In order to compare with the experiment, we subtract the two dispersions obtaining the result shown in Fig. 2b. This reproduces both the linearization induced by the interactions and a strong signal-idler emission that proves the triggering of OPO scattering.

In figure 3, real-space images of the signal polaritons are shown along with their counterpart in $k$-space. The images show that the polariton droplet undergoes unperturbed motion without diffusion in either x and y directions, proving that the dispersion around the signal state has lost its locally parabolic character. At the same time, the polariton packet also conserves a well defined momentum during all the propagation time, as seen on Figs. 3b. This confirms that the polariton signal lies on a linearized dispersion, as predicted theoretically and observed experimentally on Fig. 2. On the contrary, at low pump powers (without linearization of the dispersion), we have observed, the very fast appearance of the Rayleigh scattering circle in reciprocal space[3], and the absence of collective motion in real space (See "film_1_k_space" and "film_1_real_space", at http://www.uam.es/semicuam/films.html). This observation demonstrates the suppression of scattering as we cross the threshold into the coherent regime.

This propagation is reproduced by our numerical simulation, as seen on Fig. 4a, where a snapshot of the signal is shown at regular intervals of times. Clear propagation within the excitation spot is sustained, after the pulse has triggered the OPO (~2 ps), until the signal reaches the edge of the excitation spot. Note that the signal movement is restricted to the pump spot extension. It has a constant speed of $0.95 \times 10^6$ m s$^{-1}$, in agreement with that obtained experimentally ($1.2 \times 10^6$ m s$^{-1}$). To see the motion of the signal, we take advantage of its separation in energy from both the idler and the pump by filtering the emission in a window of energy, both in theory and experiments. In Fig. 4b, the spread of the polariton wavepacket is plotted as a function of time both from the experimental data (black points) and according to calculations based on a parabolic (red line) or linear (dashed blue) dispersions. The calculation clearly shows that the wave-packet would expand more than twice as much during the same time of flight without the linearization due to interactions. Moreover, the excellent agreement with the experiments using equation (1) implies that polariton interactions are responsible for the linearization of the signal state.

The quantum character of the polariton fluids created with a TOPO can be studied by observing their collisions against structural defects, naturally present in the sample. Figure 5-Ia shows images obtained in the near-field of a polariton fluid colliding against a defect occurring in the middle of its trajectory. In the course of its propagation, the signal shows unambiguous signs of interacting with a potential. However, it clearly maintains its cohesion in this process. This is most strikingly observed in the *k*-space counterpart, Fig. 5-Ib, that is left completely unaltered, until the very end of its trajectory, where the single-state occupancy starts spreading as the signal dies by moving off the edge of the pump laser region. Note that the images reflect the addition of two different contributions: i) that of the pump polaritons (extended in an area of $\sim 8 \times 10^3$ $\mu m^2$) and ii) that of the signal polaritons themselves, which being constantly feed by the pump reflect its density fluctuations. Figure 1c illustrates how these two contributions are detected at the signal polariton energy. The fringes observed around the defect appear due to the local change in density of the pump polaritons, which is reflected in the spatial structure of the signal. The pump polaritons are injected in a coherent state, at high energies, high density and with high *k*-vectors, with a group velocity higher than the velocity of sound[21], $v_s$, (i. e., a Mach number > 1), giving rise, in the presence of a defect, to very characteristic quantum interferences resembling Čerenkov waves, observed through the emission of the signal polaritons. Similar shockwaves have been reported recently for an atomic BEC flowing against a potential barrier at Mach numbers greater than one[9]. It is important to note that the visibility of these waves does not imply that the signal polaritons are also in the "Čerenkov" regime. On the contrary, for the signal polaritons (which are at lower energy and wavevector), the group velocity is lower than $v_s$ ($0.9\times10^6$ ms$^{-1}$, i. e. Mach < 1), and thus the signal polariton droplet can be expected and is observed to behave as a superfluid whose state is characterised by a macroscopic wavefunction with a well defined common phase; the droplet doesn't change its motion and maintains constant its wavevector while passing through the obstacle.

On Fig. 5-II, a more striking collision is observed, as the size of the defect is now comparable with that of the polariton packet. The finite-size travelling polariton fluid scatters coherently and elastically on the potential and is split in two after the collision. Note that the process is dissipationless. A normal polariton fluid[22,23] would diffuse both in real and reciprocal space in this configuration, whereas a quantum fluid whose dispersion has been linearized would pursue coherent and diffusion-less trajectories as borne out by our experiments and clearly shown in the real-space images of Fig. 5-IIa, albeit with two new momenta. The linear dispersion is the key element for this coherent propagation, as any scattered particle from the wave-packet will remain in phase with the others and at the same group velocity, precluding diffusion both in real and in reciprocal spaces. Note that this concerns the wave-packet itself, not only its excitations as is the case in the Landau picture of Helium superfluidity.

Our experimental TOPO configuration differs from the conventional realization of atomic Bose-Einstein condensates and superfluids formed spontaneously in thermal equilibrium without the action of any external driving field. In our case, due to the lossy character of polaritons, an external coherent source needs to drive the system. Furthermore, the fluid has a finite spatial extension and propagates as a whole, hence we are investigating the explicit kinematics of a droplet of BEC, rather than the indirect propagation of a defect inside an infinite size system. Also, thanks to the polariton light emission, we are able to probe continuously and in real time such features as the motion of the droplet and its dispersion that, as a manifestation of dominant interactions which strongly dress the states, is linearized when a signal exists and propagates. This dispersion is the nonlinear-response equivalent of the Bogoliubov dispersion for excitations of a superfluid. In the latter case, linearization leads to suppression of weak scattering and therefore to dissipationless motion. In our case, the dispersion reflects the dynamics of the whole system, rather than of its excitations only.

The propagation of the signal that keeps its shape unaltered over huge distances could evoke a soliton surfing on the steady state OPO, what includes correctly the notion of a non-perturbative excitation (triggered by the pulse). However, elements that are crucial for the stability of a soliton such as attractive rather than repulsive interactions, the dependence on the dimensionality of the system or on the particle densities, are not present in our case, and a plethora of different shapes, width and heights of the wavepackets can also be sustained on the same dispersion. Therefore, interactions serve the main purpose of replenishing the signal rather than holding it together, thus the soliton description is not adequate. Instead, this system provides a *coherent* and *macroscopic* population of bosons isolated in energy on a linear dispersion. Both ingredients are essential to account for the experimental findings. The system exhibits the quantum dynamics of BEC. But also in this picture, many novelties arise from the specifics of polaritons. Up to now, theoretical analysis has only been made in the perturbative regime[24] with regarding to the elementary excitation of the polariton BEC. These works show that for dissipative systems under incoherent pumping, the Bogoliubov dispersion becomes diffusive and its characteristic linearity—which is the key signature of superfluidity in the Landau picture—is lost. However, in the presence of a coherent source, like in our experiment, a truly linear mode is restored even with damping[3], and superfluid behaviour is observed. Our findings call for further work, both experimental and theoretical, to reveal new properties of this unusual state of matter: a coherent, macroscopically-occupied Bose fluid, propagating at ultra high speeds.

In conclusion, we have experimentally observed and theoretically reproduced the motion of a polariton wavepacket travelling at ultrafast velocities in a semiconductor microcavity. Along with the record breaking speed achieved by these condensed droplets, their most striking characteristics are to be found in their dynamics, characteristic of a superfluid with an unperturbed motion while crossing a weak potential and undergoing coherent scattering against strong defects. This work demonstrates that microcavity polaritons are ideal candidates to study exotic quantum bosonic phenomena. Although, as shown by this work, the novel physics associated with a quantum fluid of polaritons presents both fundamental and subtle deviations from the atomic case, these might well prove to be assets in their future studies. For instance, in a dissipative system, the particle-number conservation is lifted and a well defined phase can be externally imprinted to polaritons, allowing the investigation of symmetry breaking mechanisms.

Supplementary Videos can be found in http://www.uam.es/semicuam/films.html

ACKNOWLEDGMENTS

We thank I. Carusotto and M. Wouters for fruitful discussions and D. Steel for a critical reading of the manuscript. This work was partially supported by the Spanish MEC(MAT2005-01388, NAN2004-09109-C04-04 & QOIT-CSD2006-00019), the CAM (S-0505/ESP-0200). D.B. and E.V. acknowledge a scholarship of the FPU program of the Spanish ME. D.S and M.D.M. thank the Ramón y Cajal Program.

Correspondence and requests for materials should be addressed to D. S.: daniele.sanvitto@uam.es.

**Figures**

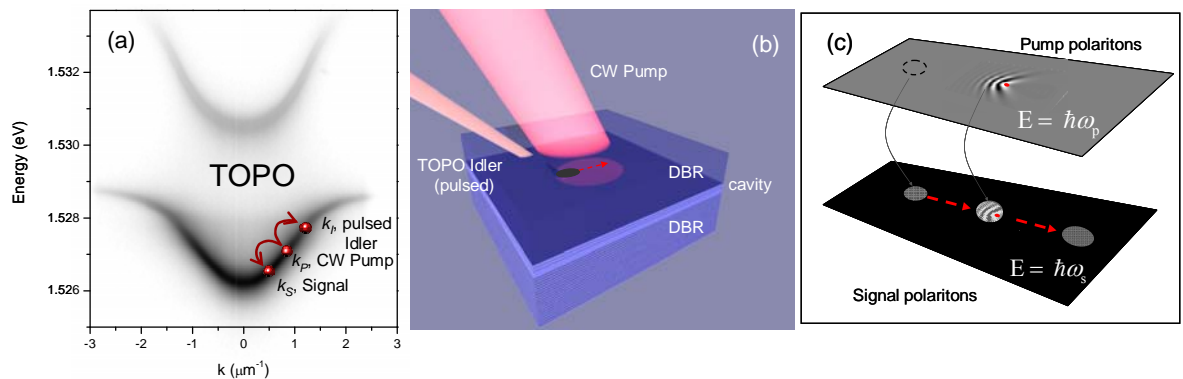

Fig. 1. (a) Experimental microcavity dispersion of the two polariton branches, showing the condition of excitation of the continuous pump and the formation of a polariton state at the CW pump energy, which feeds the signal polaritons at a lower energy, fulfilling the conditions $2k_P=k_S+k_I$ and $2E_P=E_S+E_I$. This process is initialised by a pulse at the idler state. (b) Schematic drawing of the experimental conditions. Typically, the pump injects a coherent polariton state at $10^o$ into the lower polariton branch; the pulse arrives at time t = 0 in resonance with the branch at $16^o$ triggering the OPO and generating a polariton signal with a finite momentum at 4º. The two laser beams are shown impinging on the sample surface, the polaritons are created down in the cavity region (grey circle). The upper distributed Bragg reflector (DBR) of the microcavity has been depicted as transparent for the sake of visibility. (c) Schematic representation of the motion of two fluids (pump- and signal-polaritons) against a defect depicted by the red point. Although the two fluids are spatially in the same plane, they are sketched separated in the figure to emphasize their energy difference. The lower part depicts the movement of the signal polariton fluid, at $E_S=\hbar\omega_S$, represented by the circle running from left to right on the black plane. It is possible to detect this motion thanks to the continuous feeding from the pump polaritons at $E_P = \hbar\omega_P$, which are represented in the upper part of the figure (in the gray plane).

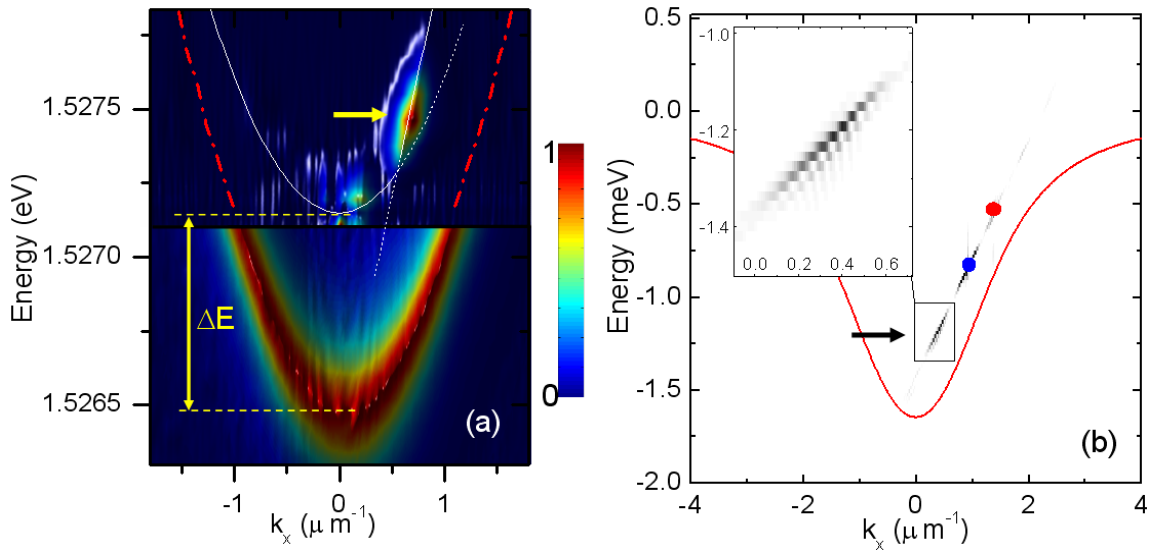

Fig. 2. (a) Upper panel: false-colour plot of the polariton dispersion. Lower panel: dispersion obtained under non-resonant low power excitation, for which polaritons are not in a coherent high density phase. The emission of upper- and lower-panels are normalised independently and shown with a linear scale. It is important to notice that the dispersion in which the signal polaritons are moving (bright spot denoted by the arrow in the upper panel) is different from the bare dispersion and shows a linear dependence on $k$ vector. The energy blue shift, $\Delta E$ (0.6 meV), is due to polariton-polariton interactions. The white parabola is plotted on top of a set of maxima of the emission too weak to be visible in the scale of the figure. (b) Computer simulation of $\left| \Psi(k_x, E) \right|^2$ showing the linearization of the signal polariton dispersion due to stimulated scattering processes occurring between a CW pump (blue circle) and a probe (red circle) after the pulse has disappeared. The ordinate is referred to the bare exciton energy. A blow up of the region around the signal is depicted in the inset. The red line describes the bare dispersion relation of polaritons.

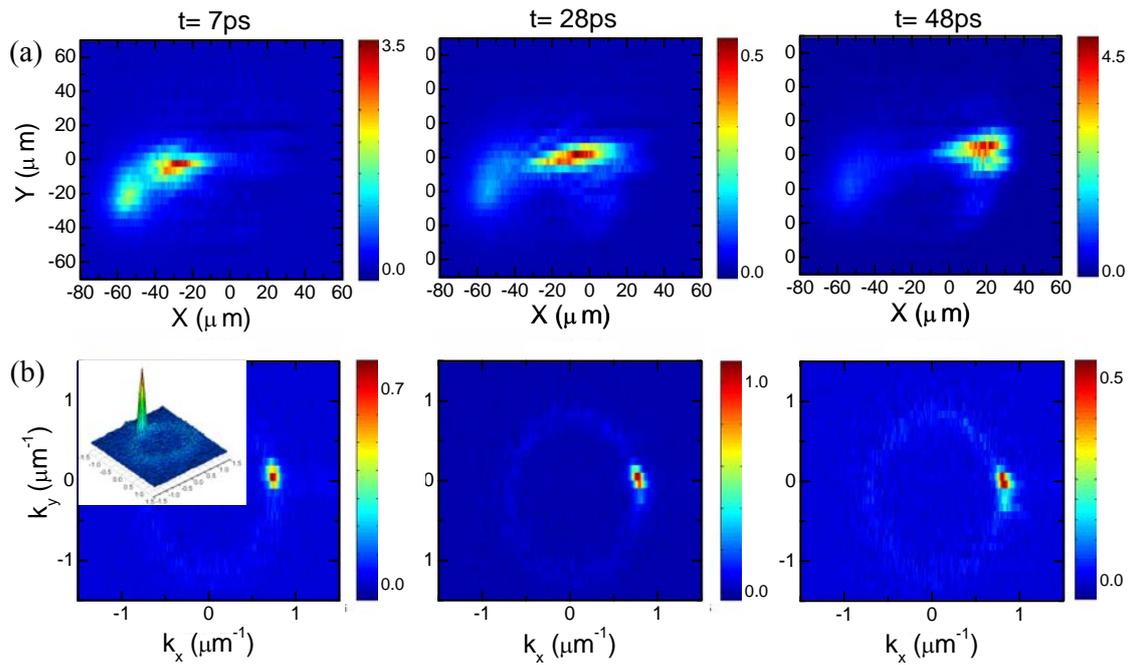

Fig. 3. (a) Spectrally selected observation at the TOPO signal energy of a coherent polariton gas moving at $v_g = 1.2 \times 10^6$ ms$^{-1}$ from left to right without being perturbed. The images, recorded under similar conditions to those of Fig. 2, are real space shots taken at different times after the pulse arrival (t = 0). The total polariton density (pump, signal and idler) is estimated to be of the order $10^2$ $\mu$m$^{-2}$. (b) Same condition as in (a) but in reciprocal (momentum) space, inset displaying a 3D view which evidence the narrow $k$ distribution. The diffusion-less motion and the invariance of the $k$-vector are a clear signature that polaritons are in a regime showing quantum coherence. Films showing polariton motion can be downloaded from http://www.uam.es/semicuam/films.html, here refer to "film_3a", and "film_3b".

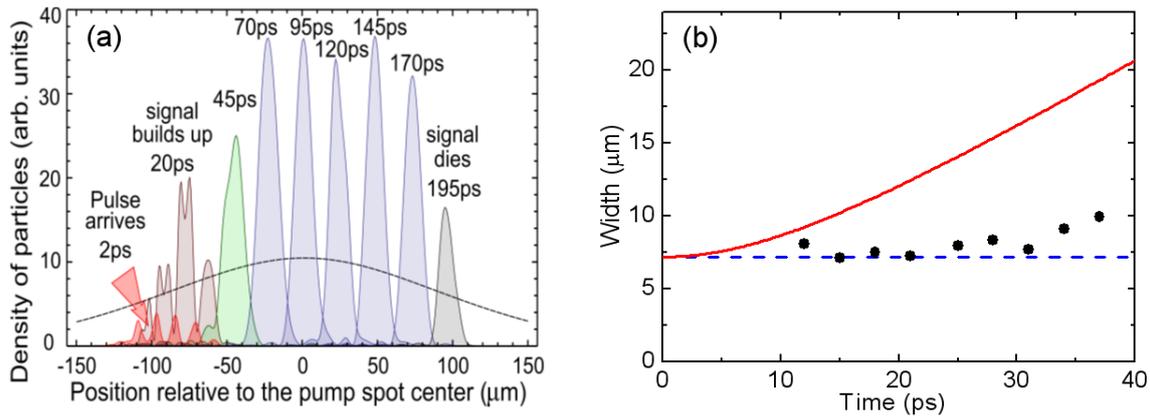

Fig. 4. (a) Computer simulation showing the spatial extent of the pump polaritons (black line) and the evolution of the signal wave-packet. The signal is created (in red) after a pulse has excited the microcavity at $k_I = 2k_P\text{-}k_S$, builds up (orange) and develops a well defined Gaussian-like wave-packet (green) that subsequently travels with small variations in shape over hundreds of microns (blue) while inside the pump-polaritons area. The wave-packet dies as it reaches the edge of the pump spot (grey). The calculation, as well as the experimental data, have been obtained by filtering in energy the emission of the microcavity at the signal polariton energy. In (b) the measured width of the coherent polariton droplet is plotted (in black points) when flowing unperturbed as a function of time. The blue dashed line shows the evolution expected for a fluid with a linear Bogoliubov-like dispersion. The red solid line depicts the size of a polariton wave-packet as obtained from the Schrödinger equation in a quadratic dispersion (diffusive regime). The animation of the simulated polariton dynamics in real and *k*-space can be downloaded from http://www.uam.es/semicuam/films.html, labelled as "film_4a".

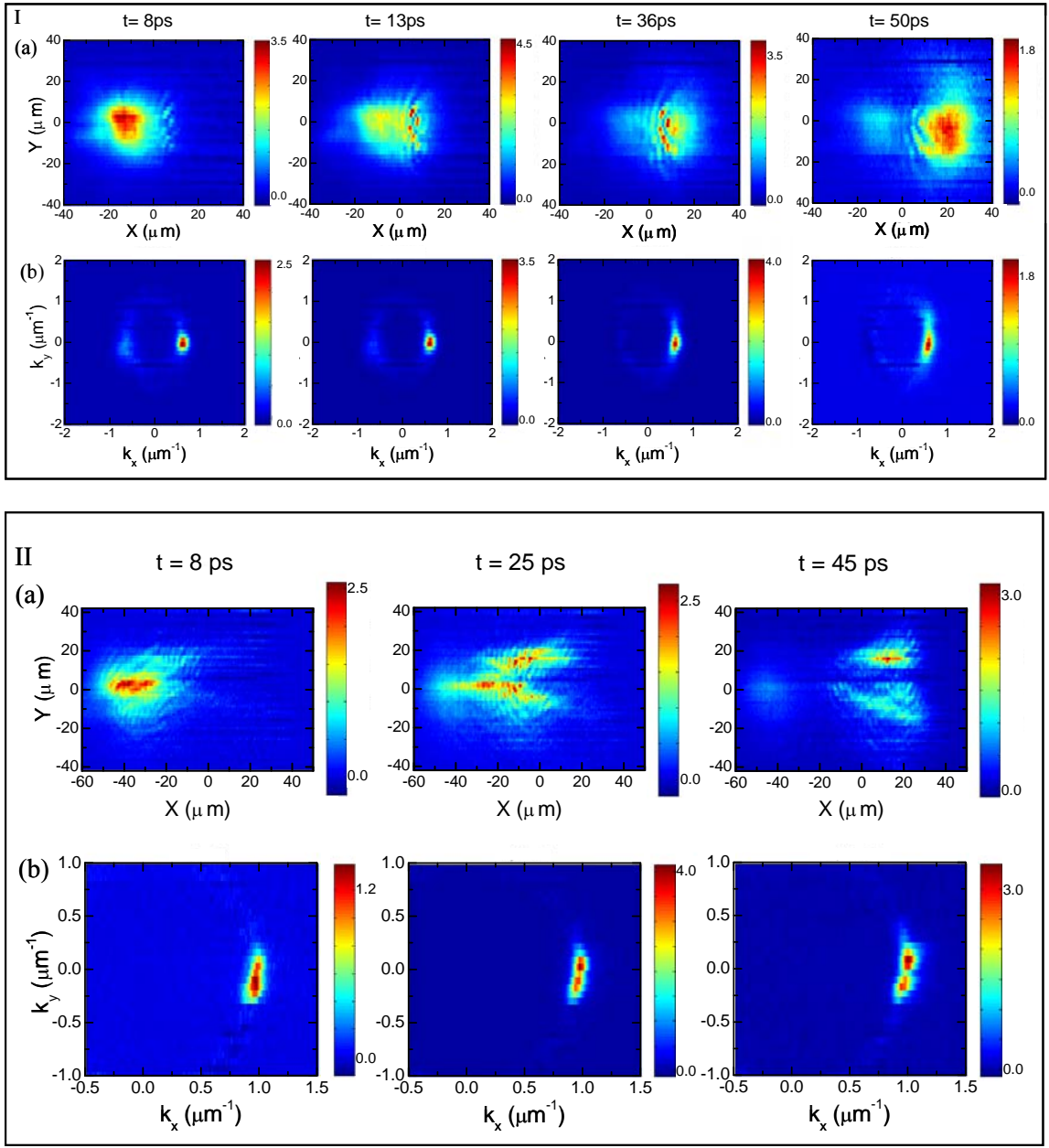

Fig. 5. (I) A small structural defect of the sample is encountered in the trajectory of the coherent polariton quantum fluid moving at a velocity of $0.9 \times 10^6$ ms$^{-1}$. The position of the obstacle is revealed, in real space (a), through Čerenkov waves caused by the pump polaritons which are travelling at a supersonic velocity. However the signal polaritons pass through the defect in a superfluid fashion without changing direction or scattering against the obstacle (see "film_5Ia" in http://www.uam.es/semicuam/films.html for a clearer proof of this phenomenon). This fact is confirmed by the images taken at the same times in momentum space (b). It is evident that the *k*-vector doesn't change significantly while the condensate runs across the obstacle (see "film_5Ib" in http://www.uam.es/semicuam/films.html). (II) In this situation the polariton superfluid is now facing a defect of size comparable with its own dimension. Under this circumstance, although still in the superfluid regime, the polariton gas is forced to "feel" the defect which is now breaking the polariton trajectory in real space (a) and showing the appearance of two independent polariton states with different *k*-vectors (b); see also "film_5IIa" and "film_5IIb" in http://www.uam.es/semicuam/films.html.

# Supplementary Information

## Quantum-fluid dynamics of microcavity polaritons

A. Amo[1], D. Sanvitto[1], D. Ballarini[1], F.P. Laussy[2], E. del Valle[2], M.D. Martin[1], A. Lemaitre[3],
J. Bloch[3], D.N. Krizhanovskii[4], M.S. Skolnick[4], C. Tejedor[2] & L. Viña[1]

[1]*Dep. Física de Materiales, Univ. Autonóma de Madrid, 28049 Madrid, Spain*
[2]*Dep. Física Teórica de la Materia Condensada, Univ. Autonóma de Madrid, 28049 Madrid, Spain.*
[3]*LPN/CNRS, Route de Nozay, 91460, Marcoussis, France*
[4]*Dep. Physics & Astronomy, Univ. of Sheffield, S3 7RH, Sheffield, U.K.*

## A Supplementary Materials and Methods

### A.1 Sample

The studied microcavity is composed by an AlAs $\lambda/2$ cavity with a top (bottom) Bragg mirror of 15 (25) $Al_{0.1}Ga_{0.9}As/AlAs$ pairs, grown on a GaAs substrate [1]. A 20 nm thick GaAs quantum well (QW) is embedded at the antinode position of the cavity mode. When the sample is kept at a temperature of 10 K, the heavy-hole excitons of the QW are in strong coupling with the cavity mode, with a Rabi splitting of 4.4 meV. The wedge shaped cavity allows fine tuning of the resonance between the QW exciton and the cavity mode by changing the position of the excitation spot on the sample.

### A.2 TOPO

Polaritons are composite bosons formed by excitons and photons. Due to the excitonic contribution, strong non-linear effects can be stimulated in microcavities. In particular, resonant excitation of the lower polariton branch (LPB) with a finite in-plane momentum gives rise to pair scattering events (similar to an Optical Parametric Oscillator, OPO) [2]. Under continuous wave pumping, a large polariton population is created at a LPB state (P) with energy $E_P$ and in-plane momentum $k_P$. If a polariton population at an idler (I) state is created by a CW probe with energy $E_I$ and momentum $k_I$, pair scattering processes will be stimulated at the signal (S) state with energy $E_S$ and momentum $k_S$, provided that the energy and momentum conservation rules can be satisfied:

$$2 \cdot E_P = E_S + E_I$$
$$2 \cdot k_P = k_S + k_I$$

This configuration is called optical parametric amplifier (OPA) [3], and is achieved in microcavities due to the particular shape of the polariton dispersion relations. In our experiment, the pump is a CW beam but the probe is a short (2 ps) pulse at the idler state, which only initializes the system, inducing a population at the S state. After the pulse has disappeared, the S state is left macroscopically occupied, and the final state stimulation of the pump polaritons to the signal polaritons carries on for hundreds of picoseconds even if the pulse is no longer present (see Fig. S1). This novel experimental configuration, only initialized by the pulse, corresponds to a triggered OPO (TOPO), where the final state stimulation of the pair scattering process is provided by the self-sustained occupancy of the S and I states and it is fuelled by the reservoir of polaritons in the P state (see Fig. 1).

## B Supplementary References

## C Supplementary Figure

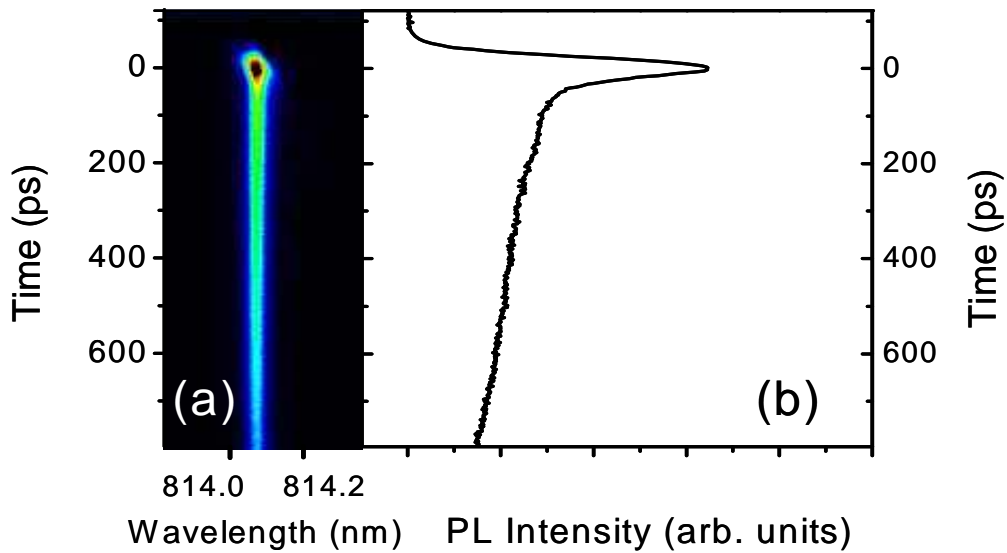

*Fig. S1: Streak camera image (a) and temporal profile (b) of the TOPO signal at k=0 after the 2 ps pulse has arrived at t=0.*

# D Supplementary Videos

Supplementary videos are available at:

http://www.uam.es/semicuam/films.html